\documentclass[preprint,prd,aps,showpacs,showkeys,nofootinbib]{revtex4}
\usepackage{graphicx}
\usepackage{dcolumn}
\usepackage{bm}
\usepackage{color}
\usepackage[dvipsnames]{xcolor}
\usepackage{amssymb}

\definecolor{light-gray}{gray}{0.78}
\definecolor{mid-gray}{gray}{0.55}
\definecolor{dark-gray}{gray}{0.32}

\begin{document}

\title{Study some two loop contribution to {muon anomalous MDM} in the N-B-LSSM }
\author{Xing-Yu Han$^{1,2,3}$, Shu-Min Zhao$^{1,2,3}$\footnote{zhaosm@hbu.edu.cn,~~hanxingyu223@163.com}, Long Ruan$^{1,2,3}$, Xi Wang$^{1,2,3}$, Xing-Xing Dong$^{1,2,3,4}$}

\affiliation{$^1$ Department of Physics, Hebei University, Baoding 071002, China}
\affiliation{$^2$ Hebei Key Laboratory of High-precision Computation and Application of Quantum Field Theory, Baoding, 071002, China}
\affiliation{$^3$ Hebei Research Center of the Basic Discipline for Computational Physics, Baoding, 071002, China}
\affiliation{$^4$ Departamento de Fisica and CFTP, Instituto Superior T$\acute{e}$cnico, Universidade de Lisboa,
Av.Rovisco Pais 1,1049-001 Lisboa, Portugal}
\date{\today}

\begin{abstract}
It is well known that the muon magnetic dipole moment (MDM) has close relation with the new physics (NP) in the development of the Standard Model (SM). Combined with the Fermilab National Accelerator Laboratory (FNAL) and the Brookhaven National Laboratory (BNL) E821 result, the departure from the SM prediction is about 5.0 $\sigma$. We study the electroweak corrections from several type two-loop SUSY diagrams and the virtual SUSY particles include chargino, neutralino, scalar lepton and scalar neutrino.
Based on the latest experimental constraints, we study the {muon anomalous MDM} under the next to the minimal supersymmetric extension of the SM with local B-L
gauge symmetry (N-B-LSSM). The abundant numerical results verify that $\tan{\beta}$, {$T_{e22}$},$~M^2_L,~M^2_e,~M_{BB'}$ play an important role in {muon anomalous MDM}. $M^2_e,~\tan{\beta}$ and {$~T_{e22}$} are sensitive parameters to {muon anomalous MDM}.
From the data obtained in all the figures of the numerical results, most of the values of  $a_{\mu}^{NBL}$ are in 2$\sigma$ interval, which can compensate the departure between the experiment data and the SM prediction.
\end{abstract}


\maketitle

\section{Introduction}

In order to further study the properties and interactions of particles, the Standard Model (SM) theory of particle physics has been gradually established and developed by Glashow, Weinberg, Salam and others \cite{b0,b1,b2,b3}. {It contains three basic  interactions of strong, weak, and electromagnetic.} However, the SM still cannot explain some physical phenomena, such as the dark energy, the dark matter, the problem of gauge hierarchy and the absence of gravity, etc. {Physicists have extended the SM based on the new symmetry of the supersymmetry (SUSY)}, resulting in the Minimal Supersymmetric Standard Model (MSSM) \cite{n0,n1,n2}. Although the MSSM can provide a dark matter candidate and alleviate hierarchy problem, it has not yet solved the $\mu$ problem and neutrino mass problem.

Based on the MSSM, next to the minimal supersymmetric extension of the SM with local B-L gauge symmetry (N-B-LSSM) extends the gauge symmetry group to {$SU(3)_C\times SU(2)_L \times U(1)_Y\times U(1)_{B-L}$}, where B represents the baryon number and L stands for the lepton number.
{In N-B-LSSM, there are three Higgs singlets and three generation right-handed neutrinos beyond MSSM.
It can produce tiny mass to light neutrinos through see-saw mechanism, and provide new dark matter candidate(light sneutrino).
}
{As the Higgs singlet $\hat{S}$ obtains a non-zero VEV ($\frac{v_S}{\sqrt{2}}$),
 the term $\lambda\hat{S}\hat{H}_u\hat{H}_d$
 can produce $\lambda\frac{v_S}{\sqrt{2}}\hat{H}_u\hat{H}_d$.
 This model does not include the $\mu$ term $\mu\hat{H}_u\hat{H}_d$,
 and $\lambda\frac{v_S}{\sqrt{2}}\hat{H}_u\hat{H}_d$ can play this role,
 which relieves the $\mu$ problem.
 Because of the introduction of three Higgs singlets,
 the neutral CP-even Higgs mass squared matrix is 5$\times$5. This can not only
 explain the 125GeV Higgs mass easily, but also enrich the Higgs physics.
Furthermore, lepton number violation and baryon number
violation processes can take place in this model, which is
 beneficial to explain the asymmetry of matter-antimatter in the universe.
With the added superfields, N-B-LSSM relieves the little hierarchy problem appearing in the MSSM.
Assuming a high scale for $v_{\eta}$, $v_{\bar{\eta}}$ and $v_S$ will suppress the corrections from some new particles.
However, their super partners are components of neutralino,
which can give considerable contributions with not very heavy mass(That is to say $M_{BL}$ and $\kappa$ are not large parameters).}

It is well known that the muon magnetic dipole moment (MDM) has close relation with the new physics (NP) in the development of the SM. The SM contributions to {muon anomalous MDM} have the following parts: 1.~the QED loop contributions \cite{g2rep2020,GWB,AKDN1,GCMH,MHBL,MDAH,AKDN2,TBPA,TAMH,GCFH,GECS,TBNC,TATK};  2.~the electroweak contributions \cite{ACWJ,CGDS}; 3.~the hadronic vacuum polarization contributions  \cite{g2rep2020,GCMH, had2}; 4.~the hadronic light-by-light contributions \cite{GCFH, GECS, TBNC}. The {muon anomalous MDM} is denoted by $a_\mu\equiv(g_\mu-2)/2$. A new result on the {muon anomalous MDM} was reported by the E989 collaboration at Fermilab \cite{Muong-2:2023cdq}:
$a_{\mu}^{FNAL}=116592055(24)\times 10^{-11}$(0.20ppm). The new averaged experiment value of muon anomaly is $a^{exp}_{\mu}=116592059(22)\times 10^{-11}$(0.35ppm). Combining all available measurements, the
SM prediction is more than 5$\sigma$ smaller than the
updated world average \cite{Datta:2023iln} : $\Delta a_\mu=a^{exp}_\mu-a^{SM}_\mu=249(48)\times 10^{-11}$.

The lattice Quantum Chromodynamics (QCD) method has been playing an increasingly important role in the precise calculation of non-perturbative low-energy hadron contributions to muon g-2, {which is an important quantity in precision tests of the SM.} In this context, the lattice QCD approach provides a first-principles framework to compute these contributions from the underlying theory of strong interactions. The lattice QCD method has the potential to reduce this deviation by providing more precise calculations of the hadronic vacuum polarization (HVP) contribution, which is a major source of uncertainty in the SM prediction for muon g-2. The data $\Delta a (HVP) = 105(59)\times 10^{-11}$ indicates that the deviation between theoretical predictions and experimental measurements will be greatly reduced. Due to the numerous experimental results, we refer to the central value of $\Delta a (HVP)$ results, which is approximately $85(56)\times  10^{-11}$ \cite{g2rep2020}. Considering the influence of QCD, it is possible to reduce the deviation between SM predictions and experiment data to 2.3 sigma.
The one-loop correction of {muon anomalous MDM} has been well studied \cite{Wang:2022wdy}, but the study of two-loop correction is more complex and not deep enough. Using the effective Lagrangian method, the authors \cite{other1} calculate and derive the leading-logarithm two-loop contributions to the {muon anomalous MDM}. The authors research corrections to {muon anomalous MDM} from the two-loop rainbow diagrams and Barr-Zee diagrams with heavy fermion sub-loop in Refs.\cite{other2,other3}. The two-loop Barr-Zee type diagrams with fermion-sub-loop and scalar-sub-loop between vector boson and Higgs are studied in BLMSSM \cite{one}. The {muon anomalous MDM} of two loop is also studied in the B-LSSM \cite{Yang:2018guw}. In this work, we study the electroweak corrections from several type two-loop SUSY diagrams and the virtual SUSY particles include chargino, neutralino, scalar lepton and scalar neutrino.

In Sec.II, we mainly introduce the
N-B-LSSM including its superpotential, the general soft breaking terms, the mass matrices and couplings. In Sec.III, we give the analytical
formulae of the one-loop and two-loop results of {muon anomalous MDM} in N-B-LSSM. The corresponding parameters and numerical analysis are shown in Sec.IV.
The last section presents our conclusions. Finally, the Appendix A shows some coupling vertices, mass matrixes and formulae that we need for this work.  {The Appendix B shows the one loop results in mass insertion approximation(MIA).}
\section{The relevant content of N-B-LSSM}

Using the local gauge group {$U(1)_{B-L}$}, we extend the MSSM to obtain the N-B-LSSM with the local gauge group {$SU(3)_C\times SU(2)_L \times U(1)_Y\times U(1)_{B-L}$}.
N-B-LSSM has new superfields beyond  MSSM, including three Higgs singlets $\hat{\chi}_1,~\hat{\chi}_2,~\hat{S}$.

\begin{table}[h]
\caption{ The superfields in N-B-LSSM}
\begin{tabular}{|c|c|c|c|c|}
\hline
Superfields & $U(1)_Y$ & $SU(2)_L$ & $SU(3)_C$ & $U(1)_{B-L}$ \\
\hline
$\hat{q}$ & 1/6 & 2 & 3 & 1/6  \\
\hline
$\hat{l}$ & -1/2 & 2 & 1 & -1/2  \\
\hline
$\hat{H}_d$ & -1/2 & 2 & 1 & 0 \\
\hline
$\hat{H}_u$ & 1/2 & 2 & 1 & 0 \\
\hline
$\hat{d}$ & 1/3 & 1 & $\bar{3}$ & -1/6  \\
\hline
$\hat{u}$ & -2/3 & 1 & $\bar{3}$ & -1/6 \\
\hline
$\hat{e}$ & 1 & 1 & 1 & $1/2$  \\
\hline
$\hat{\nu}$ & 0 & 1 & 1 & $1/2$ \\
\hline
$\hat{\chi}_1$ & 0 & 1 & 1 & -1 \\
\hline
$\hat{\chi}_2$ & 0 & 1 & 1 & 1\\
\hline
$\hat{S}$ & 0 & 1 & 1 & 0 \\
\hline
\end{tabular}
\label{quarks}
\end{table}

In the chiral superfields, $\hat H_u = \Big( {\hat H_u^ + ,\hat H_u^0} \Big)$ and $\hat H_d = \Big( {\hat H_d^0,\hat H_d^ - } \Big)$ represent the MSSM-like doublet Higgs superfields. $\hat q $ and $\hat l $ are the doublets of quark and lepton.
 $\hat u$, $\hat d$, $\hat e$ and $\hat{\nu}$ are the singlet up-type quark, down-type quark,
 charged lepton and neutrino superfields, respectively. We show the concrete forms of the two Higgs doublets and three Higgs singlets
\begin{eqnarray}
&&\hspace{-2cm}H^0_d={1\over\sqrt{2}}\phi_{d}+{1\over\sqrt{2}}v_{d}+i{1\over\sqrt{2}}\sigma_d,
\nonumber\\&&\hspace{-2cm}H^0_u={1\over\sqrt{2}}\phi_{u}+{1\over\sqrt{2}}v_{u}+i{1\over\sqrt{2}}\sigma_u,
\nonumber\\&&\hspace{-2cm}\chi_1={1\over\sqrt{2}}\phi_{1}+{1\over\sqrt{2}}v_{\eta}+i{1\over\sqrt{2}}\sigma_1,
\nonumber\\&&\hspace{-2cm}\chi_2={1\over\sqrt{2}}\phi_{2}+{1\over\sqrt{2}}v_{\bar{\eta}}+i{1\over\sqrt{2}}\sigma_2,
\nonumber\\&&\hspace{-2cm}S={1\over\sqrt{2}}\phi_S+{1\over\sqrt{2}}v_S+i{1\over\sqrt{2}}\sigma_S.
\end{eqnarray}
The vacuum expectation values(VEVs) of the Higgs superfields $H_u$, $H_d$, $\chi_1$, $\chi_2$ and $S$ are presented by
$v_u,~v_d,~v_\eta$,~ $v_{\bar\eta}$ and $v_S$ respectively. Two angles are defined as
 $\tan\beta=v_u/v_d$ and $\tan\beta_\eta=v_{\bar{\eta}}/v_{\eta}$.
\begin{eqnarray}
&&W=-Y_d\hat{d}\hat{q}\hat{H}_d-Y_e\hat{e}\hat{l}\hat{H}_d-\lambda_2\hat{S}\hat{\chi}_1\hat{\chi}_2+\lambda\hat{S}\hat{H}_u\hat{H}_d+\frac{\kappa}{3}\hat{S}\hat{S}\hat{S}+Y_u\hat{u}\hat{q}\hat{H}_u+Y_{\chi}\hat{\nu}\hat{\chi}_1\hat{\nu}
\nonumber\\&&~~~~~~~+Y_\nu\hat{\nu}\hat{l}\hat{H}_u.
\end{eqnarray}
In the superpotential for this model, $Y_{u,d,e,\nu,\chi}$ are the Yukawa couplings.
$\lambda$, $\lambda_2$ and $\kappa$ are dimensionless couplings.
  $\hat{\chi}_1,~\hat{\chi}_2,~\hat{S}$ are three Higgs singlets.  {$Y^\prime_\nu\hat{\nu}\hat{l}\hat{S}$ does not exist, because the sum of $U(1)_Y$ charges of $\hat{\nu},\hat{l},\hat{S}$ is not zero.}

The soft SUSY breaking terms are
\begin{eqnarray}
&&\mathcal{L}_{soft}=\mathcal{L}_{soft}^{MSSM}-\frac{T_\kappa}{3}S^3+\epsilon_{ij}T_{\lambda}SH_d^iH_u^j+T_{2}S\chi_1\chi_2\nonumber\\&&
-T_{\chi,ik}\chi_1\tilde{\nu}_{R,i}^{*}\tilde{\nu}_{R,k}^{*}
+\epsilon_{ij}T_{\nu,ij}H_u^i\tilde{\nu}_{R,i}^{*}\tilde{e}_{L,j}-m_{\eta}^2|\chi_1|^2-m_{\bar{\eta}}^2|\chi_2|^2\nonumber\\&&-m_S^2|S|^2-m_{\nu,ij}^2\tilde{\nu}_{R,i}^{*}\tilde{\nu}_{R,j}
-\frac{1}{2}(2M_{BB^\prime}\lambda_{\tilde{B}}\tilde{B^\prime}+\delta_{ij} M_{BL}\tilde{B^\prime}^2)+h.c~~.
\end{eqnarray}
$\mathcal{L}_{soft}^{MSSM}$ represent the soft breaking terms in the MSSM.
$T_{\kappa}$, $T_{\lambda}$, $T_2$, $T_{\chi}$ and $T_{\nu}$ are all trilinear coupling coefficients.

$U(1)_Y$ and  {$U(1)_{B-L}$} have the gauge kinetic mixing effect,
which can also be induced through RGEs even with zero value at $M_{GUT}$.
The two Abelian gauge groups are unbroken, then the basis conversion can occur with the rotation matrix $R$ ($R^T R=1$) \cite{UMSSM5,B-L1,B-L2,gaugemass}. $g_B$ is used to represent the gauge coupling constant of the  {$U(1)_{B-L}$} group.
$g_{YB}$ is used to represent the mixing gauge coupling constant of   {$U(1)_{B-L}$} group and $U(1)_Y$ group. The covariant derivatives of this model  can be written as
 {\begin{eqnarray}
&&D_\mu=\partial_\mu-i\left(\begin{array}{cc}Y,&B-L\end{array}\right)
\left(\begin{array}{cc}g_{Y},&g{'}_{{YB}}\\g{'}_{{BY}},&g{'}_{{B-L}}\end{array}\right)
\left(\begin{array}{c}B_{\mu}^{\prime Y} \\ B_{\mu}^{\prime BL}\end{array}\right)\;,
\label{gauge1}
\end{eqnarray}}
where $Y$ and $B-L$ represent the hypercharge and $B-L$ charge, respectively. The two Abelian gauge groups are unbroken, we can perform a change of basis
\begin{eqnarray}
&&\left(\begin{array}{cc}g_{Y},&g{'}_{{YB}}\\g{'}_{{BY}},&g{'}_{{B-L}}\end{array}\right)
R^T=\left(\begin{array}{cc}g_{1},&g_{{YB}}\\0,&g_{{B}}\end{array}\right)\;.
\label{gauge3}
\end{eqnarray}
As a result, the $U(1)$ gauge fields are redefined as
 {\begin{eqnarray}
&&R\left(\begin{array}{c}B_{\mu}^{\prime Y} \\ B_{\mu}^{\prime BL}\end{array}\right)
=\left(\begin{array}{c}B_{\mu}^{Y} \\ B_{\mu}^{BL}\end{array}\right)\;.
\label{gauge4}
\end{eqnarray}}
The mass matrix for neutralino in the basis $(\lambda_{\tilde{B}}, \tilde{W}^0, \tilde{H}_d^0, \tilde{H}_u^0,
\tilde{B'}, \tilde{\chi_1}, \tilde{\chi_2}, S) $ is

\begin{equation}
m_{\chi^0} = \left(
\begin{array}{cccccccc}
M_1 &0 &-\frac{1}{2}g_1 v_d &\frac{1}{2} g_1 v_u &{M}_{B B'} & 0  & 0  &0\\
0 &M_2 &\frac{1}{2} g_2 v_d  &-\frac{1}{2} g_2 v_u  &0 &0 &0 &0\\
-\frac{1}{2}g_1 v_d &\frac{1}{2} g_2 v_d  &0
&- \frac{1}{\sqrt{2}} {\lambda} v_S&-\frac{1}{2} g_{YB} v_d &0 &0 & - \frac{1}{\sqrt{2}} {\lambda} v_u\\
\frac{1}{2}g_1 v_u &-\frac{1}{2} g_2 v_u  &- \frac{1}{\sqrt{2}} {\lambda} v_S &0 &\frac{1}{2} g_{YB} v_u  &0 &0 &- \frac{1}{\sqrt{2}} {\lambda} v_d\\
{M}_{B B'} &0 &-\frac{1}{2} g_{YB} v_{d}  &\frac{1}{2} g_{YB} v_{u} &{M}_{BL} &- g_{B} v_{\eta}  &g_{B} v_{\bar{\eta}}  &0\\
0  &0 &0 &0 &- g_{B} v_{\eta}  &0 &-\frac{1}{\sqrt{2}} {\lambda}_{2} v_S  &-\frac{1}{\sqrt{2}} {\lambda}_{2} v_{\bar{\eta}} \\
0 &0 &0 &0 &g_{B} v_{\bar{\eta}}  &-\frac{1}{\sqrt{2}} {\lambda}_{2} v_S  &0 &-\frac{1}{\sqrt{2}} {\lambda}_{2} v_{\eta} \\
0 &0 & - \frac{1}{\sqrt{2}} {\lambda} v_u &- \frac{1}{\sqrt{2}}{\lambda} v_d &0 &-\frac{1}{\sqrt{2}} {\lambda}_{2} v_{\bar{\eta}}
 &-\frac{1}{\sqrt{2}} {\lambda}_{2} v_{\eta}  &\sqrt{2}\kappa v_S\end{array}
\right).\label{neutralino}
 \end{equation}

This matrix is diagonalized by the rotation matrix $N$,
\begin{equation}
N^{*}m_{\chi^0} N^{\dagger} = m^{diag}_{\chi^0}.
\end{equation}

One can find other mass matrixes in the Appendix \ref{A1}.

\section{Analytical formula}
 {\subsection{One loop results}}

\begin{figure}[h]
\setlength{\unitlength}{1mm}
\centering
\includegraphics[width=6in]{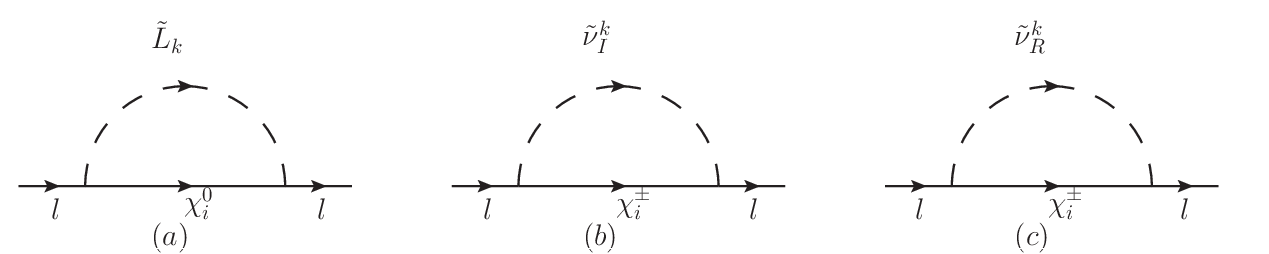}
\caption[]{ The one-loop self-energy diagrams \label{onelooptu}}
\end{figure}

With the effective Lagrangian method, the {muon anomalous MDM} can actually be expressed as
\begin{eqnarray}
&&{\cal L}_{{MDM}}={e\over4m_{l}}\;a_{l}\;\bar{l}\sigma^{\mu\nu}
l\;F_{{\mu\nu}}\label{adm},
\end{eqnarray}
here, $\sigma^{\mu\nu}=i[{\gamma}_{\mu},{\gamma}_{\nu}]/2$.$~m_{l}$ is the lepton mass, $e$ and $l$ denote the electric charge and the lepton fermion, and $F_{{\mu\nu}}$ is the electromagnetic field strength. $a_l$ is lepton MDM.

For the process $l^I\rightarrow l^I+\gamma$, {in calculating the Feynman amplitude, we use some operators defined in 6 dimensional space to describe the properties of the results}. Since the higher-dimensional operators, such as the 8-dimensional operators, are tiny, we ignore them. Their specific forms are
\begin{eqnarray}
&&\mathcal{O}_1^{L,R}=\frac{1}{(4\pi)^2}\bar{l}(i\mathcal{D}\!\!\!\slash)^3P_{L,R}l,~~~~~~~~~~~~~~
\mathcal{O}_2^{L,R}=\frac{eQ_f}{(4\pi)^2}\overline{(i\mathcal{D}_{\mu}l)}\gamma^{\mu}
F\cdot\sigma P_{L,R}l,
\nonumber\\
&&\mathcal{O}_3^{L,R}=\frac{eQ_f}{(4\pi)^2}\bar{l}F\cdot\sigma\gamma^{\mu}
P_{L,R} (i\mathcal{D}_{\mu}l),~~~~\mathcal{O}_4^{L,R}=\frac{eQ_f}{(4\pi)^2}\bar{l}(\partial^{\mu}F_{\mu\nu})\gamma^{\nu}
P_{L,R}l,\nonumber\\&&
\mathcal{O}_5^{L,R}=\frac{m_l}{(4\pi)^2}\bar{l}(i\mathcal{D}\!\!\!\slash)^2P_{L,R}l,
~~~~~~~~~~~~~~\mathcal{O}_6^{L,R}=\frac{eQ_fm_l}{(4\pi)^2}\bar{l}F\cdot\sigma
P_{L,R}l,\label{operators}
\end{eqnarray}
with $\mathcal{D}_{\mu}=\partial_{\mu}+ieA_{\mu}$ and $P_{L,R}=\frac{1\mp\gamma_5}{2}$.
The one loop contributions to the {muon anomalous MDM} are given by
\begin{eqnarray}
a_\mu^{1L}=a_{\mu}^{1L,~\tilde{L}\chi^{0}}+a_{\mu}^{1L,~\tilde{\nu}^R\chi^{\pm}}+a_{\mu}^{1L,~\tilde{\nu}^I\chi^{\pm}}.
\end{eqnarray}

The analytic form of $a_{\mu}^{1L,~\tilde{L}\chi^{0}}$, $a_{\mu}^{1L,~\tilde{\nu}^R\chi^{\pm}}$, $a_{\mu}^{1L,~\tilde{\nu}^I\chi^{\pm}}$  are as follows
\begin{eqnarray}
&&a_{\mu}^{1L,~\tilde{L}\chi^{0}}=
-\sum_{k=1}^6\sum_{i=1}^8\Big[\Re(A_L^*A_R)
\sqrt{x_{\chi_i^{0}}x_{\mu}}x_{\tilde{L}_k}\frac{\partial^2 \mathcal{G}(x_{\chi_i^{0}},x_{\tilde{L}_k})}{\partial x_{\tilde{L}_k}^2}
\nonumber\\&&\hspace{1.4cm}+\frac{1}{3}(|A_L|^2+|A_R|^2)x_{\tilde{L}_k}x_{\mu}
\frac{\partial\mathcal{G}_1(x_{\chi_i^{0}},x_{\tilde{L}_k})}{\partial x_{\tilde{L}_k}}\Big],
\nonumber\\&&a_{\mu}^{1L,~\tilde{\nu}^I\chi^{\pm}}=\sum_{i=1}^2\sum_{k=1}^6
\Big[-2\Re(B_L^{*}B_R)\sqrt{x_{\chi_i^{-}}x_\mu}\mathcal{G}_1(x_{\tilde{\nu}^I_k},x_{\chi_i^{-}})
\nonumber\\&&\hspace{1.8cm}+\frac{1}{3}(|B_L|^2+|B_R|^2)x_\mu x_{\chi_i^{-}}\frac{\partial\mathcal{G}_1(x_{\tilde{\nu}^I_k},x_{\chi_i^{-}})}{\partial x_{\chi_i^{-}}}\Big],
\nonumber\\&&a_{\mu}^{1L,~\tilde{\nu}^R\chi^{\pm}}=\sum_{i=1}^2\sum_{k=1}^6
\Big[-2\Re(C_L^{*}C_R)\sqrt{x_{\chi_i^{-}}x_\mu}\mathcal{G}_1(x_{\tilde{\nu}^R_k},x_{\chi_i^{-}})
\nonumber\\&&\hspace{1.8cm}+\frac{1}{3}(|C_L|^2+|C_R|^2)x_\mu x_{\chi_i^{-}}\frac{\partial\mathcal{G}_1(x_{\tilde{\nu}^R_k},x_{\chi_i^{-}})}{\partial x_{\chi_i^{-}}}\Big].\label{OL}
\end{eqnarray}
Here, $x=\frac{m^2}{\Lambda^2}$,$~m$ is the particle mass.  {$\Re$ represents the real part.} To save space in the text, the concrete forms of  $A_R$, $A_L$, $B_R$, $B_L$, $C_R$, $C_L$ can be found in the appendix \ref{A1}.
$\mathcal{G}(x,y)$ and $\mathcal{G}_1(x,y)$ are defined as  \cite{Zhao:2021eaa}
\begin{eqnarray}
\mathcal{G}(x,y)=\frac{1}{16 \pi
^2}\Big(\frac{x \ln x}{y-x}+\frac{y \ln
y}{x-y}\Big),~~~
\mathcal{G}_1(x,y)=(
\frac{\partial}{\partial y}+\frac{y}{2}\frac{\partial^2 }{\partial y^2})\mathcal{G}(x,y).
\end{eqnarray}

 {The one loop contributions are dominant. So, using mass insertion approximation method,
we calculate the one loop contribution to {muon anomalous MDM} in the N-B-LSSM. The specific derivation process is presented in the Appendix B.
Supposing all the masses of the superparticles are almost degenerate, we also use the following relation to obtain simplified results
\[M_1=M_2=m_{\mathcal{H}}=m_{\tilde{\mu}_L}
=m_{\tilde{\mu}_R}=m_{\tilde{\nu}^{R,I}_L}
=m_{\tilde{\nu}_R}^{R,I}=|M_{BB^\prime}|=|M_{\tilde{B}^\prime}|=M_{SUSY}.\]}
 {The simplified one loop results in N-B-LSSM are shown as
\begin{eqnarray}
&&a^{1L}_\mu\simeq \frac{1}{192\pi^2}\frac{m_\mu^2}{M_{SUSY}^2}\tan\beta(5g_2^2+g_1^2)\nonumber\\&&
\nonumber\\&&+\frac{1}{192\pi^2}\frac{m_\mu^2}{M_{SUSY}^2}\tan\beta\texttt{sign}[M_{\tilde{B}^\prime}](g_B^2+3g_{YB}g_B+g_{YB}^2)
\nonumber\\&&+\frac{1}{960\pi^2}\frac{m_\mu^2}{M_{SUSY}^2}\tan\beta
g_1(4g_{YB}+3g_B)\texttt{sign}[M_{BB^\prime}]
\Big(1-4\texttt{sign}[M_{\tilde{B}^\prime}]\Big).\label{amuS1}
\end{eqnarray}
The first line in Eq.(\ref{amuS1}) is the MSSM one loop results, which
  increase with the enlarging $\tan\beta$.
   It indicates that large $\tan\beta$ leads to large {muon anomalous MDM} in MSSM.}

 {The results in the second and
third lines of Eq.(\ref{amuS1}) correspond to the N-B-LSSM
contribution beyond MSSM,  and the values of the
parameters $M_{\tilde{B}^\prime}$, $M_{BB'}$, $g_{YB}$
can be positive or negative. $g_B$ is always positive.
Therefore, the sum of the second and third lines of Eq.(\ref{amuS1}) can be negative.
It implies that large $\tan\beta$ can lead to small MDM results.}
{When $M_{\tilde{B}^\prime}$ is negative and $M_{BB'}$ is also negative, the results decreases with increasing $\tan{\beta}$. This characteristic can be embodied clearly by the following formula}
{\begin{eqnarray}
&&a^{1L}_\mu\simeq \frac{1}{192\pi^2}\frac{m_\mu^2}{M_{SUSY}^2}\tan\beta(5g_2^2+g_1^2)
-\frac{1}{192\pi^2}\frac{m_\mu^2}{M_{SUSY}^2}\tan\beta(g_B^2+3g_{YB}g_B+g_{YB}^2)
\nonumber\\&& ~~~~~~~-\frac{1}{192\pi^2}\frac{m_\mu^2}{M_{SUSY}^2}\tan\beta
g_1(4g_{YB}+3g_B).\label{amuS1}
\end{eqnarray}}

We ignore the contributions of neutral Higgs-lepton and charged Higgs-neutrino, which are suppressed by the square of the Higgs-lepton coupling $\frac{m_{\mu}^2}{m^2_W}\sim10^{- 6}$.
We neglect the one-loop contribution of $M_{Z^\prime}$-muon. Since the mass of the new vector boson $M_{Z^\prime}$ is greater than 5.1 TeV \cite{at,gc}, the one-loop contribution of $M_{Z^\prime}$-muon is suppressed by the factor $\frac{m_Z^2}{m^2_{Z^\prime}}\sim 4\times 10^{-4}$.

\begin{figure}[h]
\setlength{\unitlength}{1mm}
\centering
\includegraphics[width=6in]{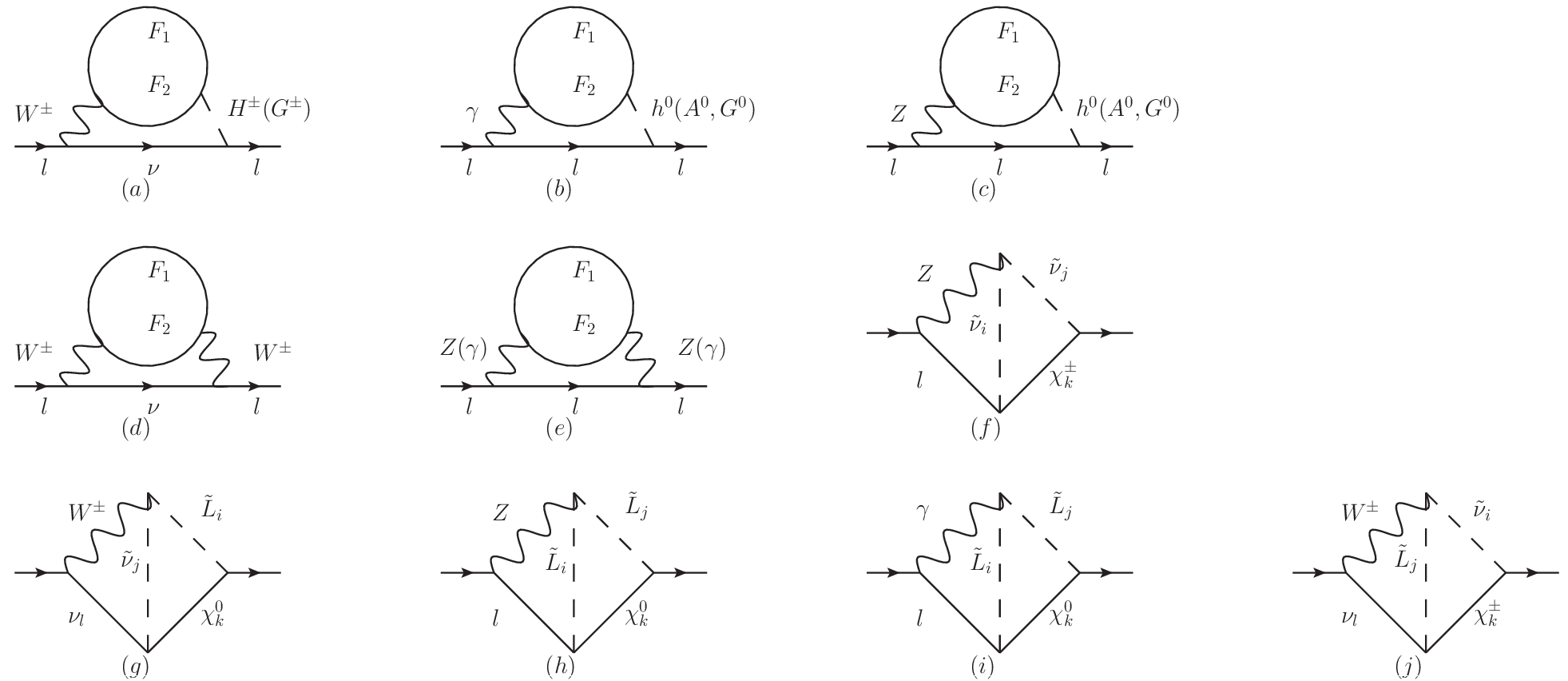}
\caption[]{ The two-loop self-energy diagrams \label{twolooptu}}
\end{figure}

 {\subsection{Two loop results}}

The main contributions of two-loop graphs to {muon anomalous MDM} come from the following

1. The two-loop Barr-Zee type diagrams (Fig.\ref{twolooptu}(a), Fig.\ref{twolooptu}(b) and Fig.\ref{twolooptu}(c)) with fermion sub-loop. In the work, the authors focus on their contributions to {muon anomalous MDM}\cite{other2}. On the supposition $\chi^\pm\sim\chi^0\sim M$, {we can get approximate results with the factor}
$\frac{x_\mu}{x_M^{1/2}x_V^{1/2}}=\frac{m^2_\mu}{Mm_V}$.
$m_V$ represents the mass of heavy vector bosons $m_Z\sim m_W\sim m_V$. This analysis is in the mass eigenstate,
and overall the rotation matrices to diagonalize the particle mass matrices should be taken into account. Then the order analysis
in the mass insertion approximation is more appreciated, which shows the order as $\frac{m_\mu^2}{M^2}\tan\beta$.

2. Fig.\ref{twolooptu}(d) and Fig.\ref{twolooptu}(e) are the two-loop rainbow diagrams with fermion sub-loop and the vector bosons ($\gamma$, $Z$, $W$). They have important contributions to {muon anomalous MDM}\cite{ffa,lepton}.

3. The two-loop self-energy diagrams (Fig.\ref{twolooptu}(f), $\dots$, Fig.\ref{twolooptu}(j)) belong to the diamond type. The diamond type diagrams in Ref.\cite{two2,our} possess large factors.
This type two-loop diagrams studied in this work contains five virtual particles including: one vector boson, two scalars and two fermions.

With the assumption $m_{F_1}=m_{F_2}\gg m_W$, the results \cite{ffa} for the Fig.\ref{twolooptu} (a),(b),(c) can be simplified as
\begin{eqnarray}
&&a_\mu^{2L,~WH}=\frac{eC_{\bar{\mu}H\nu}^L}{512\sqrt{2}\pi^4s_W}\sum_{F_1=\chi^{\pm}}\sum_{F_2=\chi^0}\frac{x_\mu^{1/2}}{ x^{1/2}_{F_1}}\Big\{\frac{199}{36}\Re(C_{H\bar{F}_1F_2}^LC_{W\bar{F}_2F_1}^L+H_{C\bar{F}_1F_2}^RC_{W\bar{F}_2F_1}^R)
\nonumber\\&&\hspace{1.6cm}+\Big[\frac{13}{3}+2
(\ln{x_{F_1}}-\varrho_{1,1}(x_W,x_{H^\pm}))\Big]\Re(C_{H\bar{F}_1F_2}^LC_{W\bar{F}_2F_1}^R+C_{H\bar{F}_1F_2}^RC_{W\bar{F}_2F_1}^L)
\nonumber\\
&&\hspace{1.6cm}+\Big[\frac{4}{3}(\ln{x_{F_1}}-\varrho_{1,1}(x_W,x_{H^\pm}))-\frac{16}{9}\Big]\Re(C_{H\bar{F}_1F_2}^LC_{W\bar{F}_2F_1}^L\hspace{-0.1cm}
-\hspace{-0.1cm}C_{H\bar{F}_1F_2}^RC_{W\bar{F}_2F_1}^R)
\nonumber\\
&&\hspace{1.6cm}+\Big[\frac{2}{9}-\frac{8}{3}(\ln{x_{F_1}}-\varrho_{1,1}(x_W,x_{H^\pm}))\Big]\Re(C_{H\bar{F}_1F_2}^LC_{W\bar{F}_2F_1}^R\hspace{-0.1cm}
-\hspace{-0.1cm}C_{H\bar{F}_1F_2}^RC_{W\bar{F}_2F_1}^L)\Big\},
\\
&&a_\mu^{2L,~WG}
=\frac{eC_{\bar{\mu}G\nu}^L}{512\sqrt{2}\pi^4s_W}\sum_{F_1=\chi^{\pm}}\sum_{F_2=\chi^0}\frac{x_\mu^{1/2}}{ x^{1/2}_{F_1}}\Big\{\frac{199}{36}\Re(C_{G\bar{F}_1F_2}^LC_{W\bar{F}_2F_1}^L+C_{G\bar{F}_1F_2}^RC_{W\bar{F}_2F_1}^R)
\nonumber\\&&\hspace{1.6cm}+\Big[\frac{7}{3}+2
(\ln{x_{F_1}}-\ln x_W)\Big]\Re(C_{G\bar{F}_1F_2}^LC_{W\bar{F}_2F_1}^R+C_{G\bar{F}_1F_2}^RC_{W\bar{F}_2F_1}^L)
\nonumber\\
&&\hspace{1.6cm}+\Big[\frac{4}{3}(\ln{x_{F_1}}-\ln x_W)-\frac{28}{9}\Big]\Re(C_{G\bar{F}_1F_2}^LC_{W\bar{F}_2F_1}^L\hspace{-0.1cm}
-\hspace{-0.1cm}C_{G\bar{F}_1F_2}^RC_{W\bar{F}_2F_1}^R)
\nonumber\\
&&\hspace{1.6cm}+\Big[\frac{26}{9}-\frac{8}{3}(\ln{x_{F_1}}-\ln x_W)\Big]\Re(C_{G\bar{F}_1F_2}^LC_{W\bar{F}_2F_1}^R\hspace{-0.1cm}
-\hspace{-0.1cm}C_{G\bar{F}_1F_2}^RC_{W\bar{F}_2F_1}^L)\Big\}.\label{awg}
\\
&&a_\mu^{2L,~\gamma h^0}=\frac{e^2}{64\sqrt{2}\pi^4}C_{h^0\bar{\mu}\mu}\sum_{F_1=F_2=\chi^\pm}\frac{x_\mu^{1/2}}{ x^{1/2}_{F_1}}
\Re(C_{h^0\bar{F}_1F_2}^L)\Big[1+\ln\frac{x_{F_1}}{x_{h^0}}\Big],
\\
&&a_\mu^{2L,~Zh^0}=\frac{\sqrt{2}}{512\pi^4}\sum_{F_1=F_2=\chi^{\pm},\chi^0}
C_{h^0\bar{\mu}\mu}\frac{x_\mu^{1/2}}{ x^{1/2}_{F_1}}\Big[\varrho_{1,1}(x_Z,x_{h^0})-\ln{x_{F_1}}-1\Big]\times(C^L_{Z\bar{\mu}\mu}+C^R_{Z\bar{\mu}\mu})
\nonumber\\&&\hspace{1.6cm}\Re(C_{h^0\bar{F}_1F_2}^LC_{Z\bar{F}_2F_1}^L+C_{h^0\bar{F}_1F_2}^RC_{Z\bar{F}_2F_1}^R).
\end{eqnarray}

To save space in the text, the complete calculation process and the results of other two-loop diagrams can be found in our previous work \cite{Zhao:2021eaa}. The corrections to {muon anomalous MDM} from the studied two-loop diagrams are
\begin{eqnarray}
&&\hspace{1.0cm}a_\mu^{2L}=a_\mu^{2L,~BZ}+a_\mu^{2L,~RB}+a_\mu^{2L,~DIA},
\nonumber\\&&\hspace{1.0cm}a_\mu^{2L,~BZ}=a_\mu^{2L,~WH}+a_\mu^{2L,~WG}+a_\mu^{2L,~\gamma h_0}+a_\mu^{2L,~\gamma G_0}+a_\mu^{2L,~\gamma A_0}\nonumber\\&&\hspace{2.5cm}+a_\mu^{2L,~Z h_0}+a_\mu^{2L,~Z G_0}+a_\mu^{2L,~Z A_0},
\nonumber\\&&\hspace{1.0cm}a_\mu^{2L,~RB}=a_{\mu}^{2L,~WW}
+a_{\mu}^{2L,~ZZ}+a_{\mu}^{2L,~Z\gamma}+a_{\mu}^{2L,~\gamma\gamma},
\nonumber\\&&\hspace{1.0cm}a_\mu^{2L,~DIA}=a^{2L,~Z\tilde{\nu}\chi^{\pm}}_{\mu}+a^{2L,~Z\tilde{L}\chi^{0}}_{\mu}+a^{2L,~\gamma\tilde{L}\chi^{0}}_{\mu}
+a^{2L,~W\tilde{L}\tilde{\nu}\chi^{0}}_{\mu}+a^{2L,~W\tilde{L}\tilde{\nu}\chi^{-}}_{\mu}.
\end{eqnarray}
The concrete expressions can be found in Appendix \ref{A1}.

At two-loop level, including the one-loop results and two-loop results, the {muon anomalous MDM} is given by
\begin{eqnarray}
&& a_{\mu}^{NBL}=a_{\mu}^{1L}+a_{\mu}^{2L}.
\end{eqnarray}
\section{Numerical analysis}
To take the numerical calculation, some restrictions should be taken into account.

1. The experimental value of $\tan{\beta}_{\eta}$ should be less than 1.5 in order to meet the LHC experimental data \cite{lb,ATLAS:2014pcp,CMS:2014lcz}.

2. We consider the experimental constraints from the lightest CP-even Higgs $h^0$ mass is around 125.25 GeV \cite{pdg2022,higgs}.

3. The $Z^\prime$ boson mass is larger than 5.1 TeV. The ratio between $M_{Z^\prime}$ and its gauge $M_{Z^\prime}/g_B \geq 6 ~{\rm TeV}$ \cite{at,gc}.

4. For particles that exceed the SM, the mass limits considered are: the slepton mass is greater than $700~{\rm GeV}$, and the chargino mass is greater than $1100 ~{\rm GeV}$ \cite{pdg2022}.

{5. The limitation of Charge and Color Breaking(CCB) \cite{ccb1,ccb2} is considered.
}

Considering these limitations, we adopt the following parameters:
\begin{eqnarray}
&&~~\kappa=0.1,~\tan{\beta}_{\eta}=0.9,~g_{B}=0.3, {~v_{B-L}=17~{\rm TeV},}~v_S=4~{\rm TeV},
\nonumber\\&&~~T_{\lambda} =~T_{\lambda_2} =1~ {\rm TeV},~T_{uii}=1~{\rm TeV},{~T_{e11}=~T_{e33}=10~ {\rm GeV}},
\nonumber\\&&~~\lambda=0.4,~~M_1=0.1~{\rm TeV},~M_2=1.2~{\rm TeV},~~\lambda_2=-0.25,~T_{\kappa} =-2.5~ {\rm TeV},
\nonumber\\&&~~Y_{\nu_{11}} = 1.09285\times10^{-6},~~Y_{\nu_{22}} = 1.4\times10^{-6},~~Y_{\nu_{33}} = 1.35242\times10^{-6},
\nonumber\\&&~~Y_{\nu_{12}} = 7.6042\times10^{-8},~~Y_{\nu_{13}}= 4.51693\times10^{-8},~~Y_{\nu_{23}} = 2.80323\times10^{-7}.
\end{eqnarray}

We generally take the values of new particle masses($M_{BB'}$, $M_{BL}$) near the order of $10^{3}$ GeV, which is around the energy scale of new physics. $T_{\lambda}$ and $T_{2}$ etc. are trilinear coupling coefficients, which are roughly in the order of magnitude of the mass, and can be varied up or down to the order of $10^{2} \sim 10^{4}$ GeV. $M^2_L$, $M^2_e$ are all of mass square dimension, and can be up to the order of $10^{6}~ {\rm GeV}^2$. The dimensionless parameters $\lambda$ and $\lambda_2$ etc. are generally taken as numbers less than 1.

In the following numerical analysis process, the parameters that need to be studied are:
\begin{eqnarray}
&&~\tan{\beta},~~M^2_{lii}=M^2_L,~~M^2_{eii}=M^2_e,
\nonumber\\&&~~M_{BB'},{~T_{e22}},~~M_{BL},~~g_{YB}~~(i=1,2,3).
\end{eqnarray}
In addition to the above parameters, non diagonal elements are defined as zero.

\subsection{The one-dimensional graphs}
We use the simplified expression of $\chi^2_{NBL}$ as
\begin{eqnarray}
\chi^2_{NBL}=e^{-\left(\frac{\Delta a_\mu- a_{\mu}^{NBL}}{\delta_ {a_\mu}}\right)^2},
\end{eqnarray}
with $\Delta a_\mu=249\times 10^{-11}$, $\delta_ {a_\mu}=48\times 10^{-11}$. This formula clearly indicates the deviation of $a_{\mu}^{NBL}$ and $\Delta a_\mu$. When $a_{\mu}^{NBL}$ approaches $\Delta a_\mu$, $\chi^2_{NBL}$ approaches 1.

The light gray and light orange regions in all figures represent the experimental limits of $\Delta a_\mu$, where the light gray region represents the 1$\sigma$ range of $\Delta a_\mu$, and the light orange region represents the 2$\sigma$ range of $\Delta a_\mu$. {In figures 3,4,5, the orange line represents that $a_{\mu}^{NBL}$ corresponds to the left ordinate axis, and the blue line represents that $\chi^2_{NBL}$ corresponds to the right ordinate axis.}

\begin{figure}[ht]
\setlength{\unitlength}{5mm}
\centering
\includegraphics[width=3.15in]{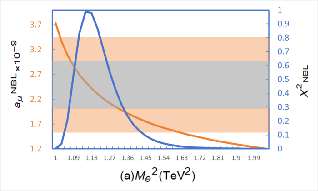}
\vspace{0.2cm}
\setlength{\unitlength}{5mm}
\centering
\includegraphics[width=3.15in]{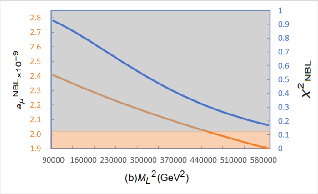}
\caption{$a_{\mu}^{NBL}$~and~$\chi^2_{NBL}$ in $M^2_e$ plane (a) and $M^2_L$ plane (b).}{\label {D1}}
\end{figure}

With the parameters $\tan{\beta}=10$,{$~T_{e22}=2.4~{\rm TeV}$},$~M_{BB'}=0.1~{\rm TeV}$ and$~g_{YB}=0.1$ in Fig.\ref{D1}(a), we plot $M^2_e$ versus $a_{\mu}^{NBL}$~and~$\chi^2_{NBL}$. {The $a_{\mu}^{NBL}$ is decreasing function as $M^2_e$ turns large in the range of $1~{\rm TeV}^2 < M^2_e < 2~{\rm TeV}^2$. When $M^2_e$ is in the range of $1.09~{\rm TeV}^2$ to $1.29~{\rm TeV}^2$, $a_{\mu}^{NBL}$ is in the 1$\sigma$ interval. When $M^2_e$ is in the range of $1~{\rm TeV}^2$ to $1.09~{\rm TeV}^2$ and $1.29~{\rm TeV}^2$ to $2~{\rm TeV}^2$,  $a_{\mu}^{NBL}$ is in the 2$\sigma$ interval or larger interval}. The expression of $\chi^2_{NBL}$ indicates that as the value of $a_{\mu}^{NBL}$ deviates from $\Delta a_\mu$, $\chi^2_{NBL}$ becomes smaller. When $a_{\mu}^{NBL}$ approaches $\Delta a_\mu$, $\chi^2_{NBL}$ approaches 1. The image meanings of $\chi^2_{NBL}$ and $a_{\mu}^{NBL}$ in the figure are consistent. We plot $M^2_L$ versus $a_{\mu}^{NBL}$~and~$\chi^2_{NBL}$ in the Fig.\ref{D1}(b). $a_{\mu}^{NBL}$~and~$\chi^2_{NBL}$ show a downward trend with the increase of $M^2_L$. $a_{\mu}^{NBL}$ is mostly in the range of 1$\sigma$, and the trend of $\chi^2_{NBL}$ is consistent with $a_{\mu}^{NBL}$.

 $M^2_L$ is the parameter appearing in the mass matrices of the CP-odd sneutrino,
the CP-even sneutrino, and the slepton. $M^2_e$ just appears in the mass matrix of the slepton.
The increase of $M^2_e$ and $M^2_L$ makes sneutrino and slepton heavy, which suppresses  the contributions from the CP-odd sneutrino,
the CP-even sneutrino, and the slepton. Thereby  $a_{\mu}^{NBL}$  decreases with the increase of $M^2_e$ and $M^2_L$.

\begin{figure}[ht]
\setlength{\unitlength}{5mm}
\centering
\includegraphics[width=3.15in]{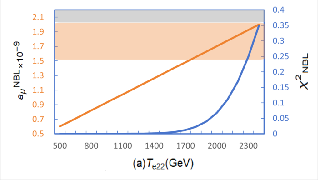}
\vspace{0.2cm}
\setlength{\unitlength}{5mm}
\centering
\includegraphics[width=3.15in]{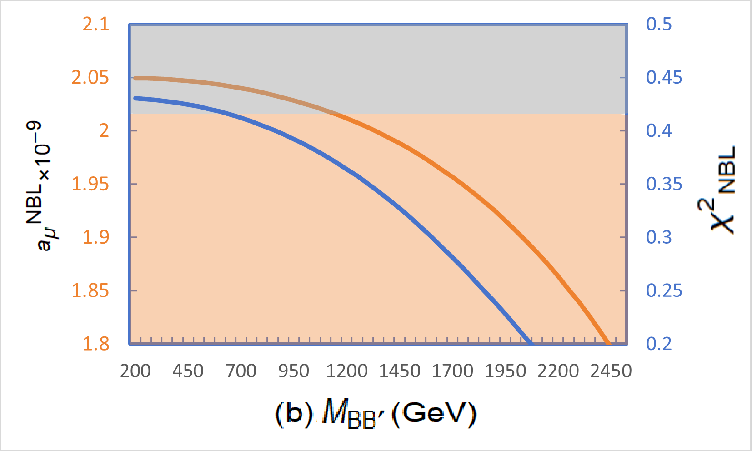}
\caption{$a_{\mu}^{NBL}$~and~$\chi^2_{NBL}$ in $T_{e22}$ plane (a) and $M_{BB'}$ plane (b).}{\label {D2}}
\end{figure}

We suppose $\tan{\beta}=10$,$~M^2_e=1.5~{\rm TeV^2}$,$~M^2_L=0.16~{\rm TeV^2}$,$~g_{YB}=0.1$. Similarly, we plot {$T_{e22}$} versus $a_{\mu}^{NBL}$~and~$\chi^2_{NBL}$ in the Fig.\ref{D2}(a). In this figure, $a^{NBL}_{\mu}$ significantly increases with the growth of {$T_{e22}$}. When {$T_{e22}$} is greater than $1.7~{\rm TeV}$, it enters the 2$\sigma$ range. $\chi^2_{NBL}$ is also showing a growth trend. In addition, we study
the parameter $M_{BB'}$ influences on $a_{\mu}^{NBL}$~and~$\chi^2_{NBL}$ in Fig.\ref{D2}(b). When $0.2~{\rm TeV} < M_{BB'} < 2.45~{\rm TeV}$, both lines decrease with the increase of $M_{BB'}$. {When $0.2~{\rm TeV} < M_{BB'} < 1.2~{\rm TeV}$, $a_{\mu}^{NBL}$ is in the 1$\sigma$ interval. When $M_{BB'}$ is in the range of $1.2~{\rm TeV}^2$ to $2.45~{\rm TeV}^2$, $a_{\mu}^{NBL}$ is in the 2$\sigma$ interval.} $M_{BB'}$ is the mass of the two U(1) gauginos mixing, and appears as the non-diagonal element of the neutralino mass matrix. The increase in $M_{BB'}$ has affected the neutralino mass matrix, leading to a downward trend in $a^{NBL}_{\mu}$, and the curve of $\chi^2_{NBL}$ indicates that this effect is relatively strong.

\begin{figure}[ht]
\setlength{\unitlength}{5mm}
\centering
\includegraphics[width=3.15in]{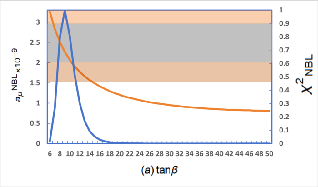}
\vspace{0.2cm}
\setlength{\unitlength}{5mm}
\centering
\includegraphics[width=3in]{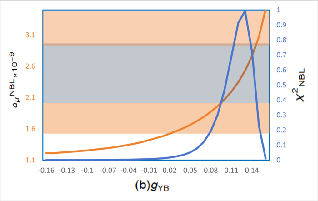}
\caption{$a_{\mu}^{NBL}$~and~$\chi^2_{NBL}$ in $\tan{\beta}$ plane (a) and $g_{YB}$ plane (b).}{\label {D3}}
\end{figure}

We use the parameters as $~M_{BB'}=0.1~{\rm TeV}$,$~M^2_e=1.5~{\rm TeV^2}$,$~M^2_L=0.16~{\rm TeV^2}$,{$~T_{e22}=2.4~{\rm TeV}$}  in Fig.\ref{D3}.
Next, we plot $\tan{\beta}$ versus $a_{\mu}^{NBL}$~and~$\chi^2_{NBL}$ in the Fig.\ref{D3}(a). The $a_{\mu}^{NBL}$ is decreasing function as $\tan{\beta}$ turns large in the range of {$6 <\tan{\beta} < 50$}. {The downward trend of $a^{NBL}_{\mu}$ is relatively gentle, when $\tan{\beta}$ is greater than 20.} The downward trend of $\chi^2_{NBL}$ is relatively severe. $\tan\beta$ must be a sensitive parameter because it appears almost in all mass matrices of fermions, scalars, and Majoranas, and it can affect the vertex couplings and masses of particles by directly affecting $v_d$ and $v_u$. By influencing these factors, $\tan{\beta}$ shows a downward trend.

Finally, we plot $g_{YB}$ versus $a_{\mu}^{NBL}$~and~$\chi^2_{NBL}$ in the Fig.\ref{D3}(b). $g_{YB}$ is the gauge kinetic mixing coupling constant which arises from the existence of two Abelian gauge groups. In the figure, $a^{NBL}_{\mu}$ increases with the increase of $g_{YB}$, { $\chi^2_{NBL}$ increases first and then decreases with the increase of $g_{YB}$}. When $g_{YB}$ is greater than {0.02}, it enters the 2$\sigma$ range, and when $g_{YB}$ is greater than {0.09}, it enters the 1$\sigma$ range. The $g_{YB}$ might influence the contributions of the slepton and  other particles and the corresponding couplings, thereby causing $a^{NBL}_{\mu}$ to increase with the increase of $g_{YB}$.

\subsection{The multi-dimensional scatter plots}

In this section of the work, we select six parameters $\tan{\beta}$,{$~T_{e22}$},$~M^2_{L}$,$~M_{BL}$,$~M_{BB'}$ and $M^2_{e}$ discussed in the one-dimensional graph to draw the scatter plots.

\begin{table}[ht]
\caption{ The meaning of shape style in Fig. \ref{D4} and Fig. \ref{D5}}
\begin{tabular}{|c|c|}
\hline
Shape style&Fig. \ref{D4} and Fig. \ref{D5}\\
\hline
\textcolor{blue}{$\blacktriangle$} & $0 < a_{\mu}^{NBL}<2.01\times10^{-9}$\\
\hline
\textcolor{red}{$\blacksquare$} & $2.01\times10^{-9}\leqslant a_{\mu}^{NBL}<2.97\times10^{-9}$ \\
\hline
\textcolor{Green}{$\bullet$}&$2.97\times10^{-9}\leqslant a_{\mu}^{NBL}$  \\
\hline
\end{tabular}
\label{t12}
\end{table}

\begin{figure}[ht]
\setlength{\unitlength}{5mm}
\centering
\includegraphics[width=3.2in]{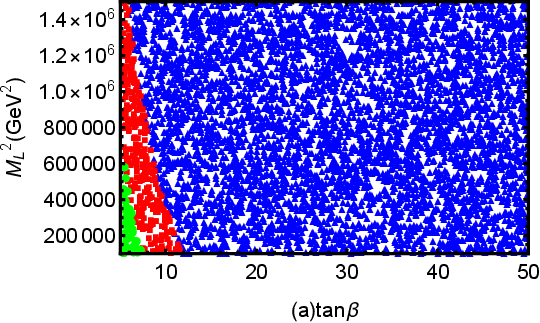}
\vspace{0.2cm}
\setlength{\unitlength}{5mm}
\centering
\includegraphics[width=3.1in]{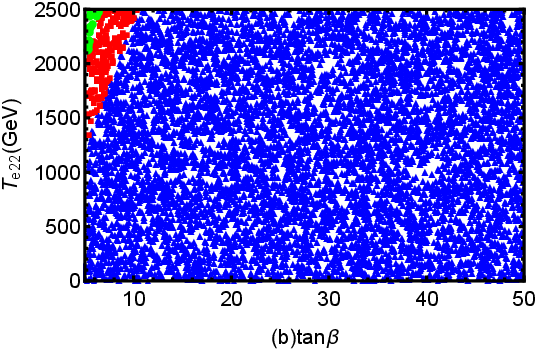}
\caption{$a_{\mu}^{NBL}$ in $\tan{\beta}-M^2_L$ plane(a), $\tan{\beta}-T_{e22}$ plane(b).}{\label {D4}}
\end{figure}

Supposing {$~T_{e22}=2.4~{\rm TeV}$},$~M^2_e=1.5~{\rm TeV^2}$,$~M_{BB'}=0.1~{\rm TeV}$,{$~M_{BL}=1~{\rm TeV}$}, we plot $a_{\mu}^{NBL}$ in the $\tan{\beta}$ versus $M^2_L$  in the Fig.\ref{D4}(a). Within the range of $ 5 < \tan{\beta} < 50$ and {$ 0.1~{\rm TeV^2} < M^2_L < 1.4~{\rm TeV^2}$,}
the \textcolor{Green}{$\bullet$} and  \textcolor{red}{$\blacksquare$} are slightly tilted and distributed on the left side of the graph. When $ 5 < \tan{\beta} < 7$, the proportion of \textcolor{Green}{$\bullet$} is higher. When $ 7 < \tan{\beta} < 13$, the proportion of  \textcolor{red}{$\blacksquare$} is higher. When $ 13 < \tan{\beta} < 50$, there are  only \textcolor{blue}{$\blacktriangle$}. We can conclude that $a_{\mu}^{NBL}$ decreases with the increase of $\tan{\beta}$, and the trend is more intense. $a_{\mu}^{NBL}$ also decreases with the increase of $M^2_L$, but the trend is relatively gentle. This is consistent with the decreasing trend of $a_{\mu}^{NBL}$ as $\tan{\beta}$ and $M^2_L$, which are shown in Fig.\ref{D1}(b) and Fig.\ref{D3}(a). And the reduction amplitude is also consistent with the one-dimensional graph.

With $~M^2_L=0.16~{\rm TeV^2}$,$~M^2_e=1.5~{\rm TeV^2}$,$~M_{BB'}=0.1~{\rm TeV}$,{$~M_{BL}=1~{\rm TeV}$}, Fig.\ref{D4}(b) displays a plot of $a_{\mu}^{NBL}$ in the $\tan{\beta}$ versus {$T_{e22}$} plane. We can clearly see that the space is roughly divided into three parts. The \textcolor{Green}{$\bullet$} present a triangular shape distributed in the upper left, the  \textcolor{red}{$\blacksquare$} present a strip close to the green points, and the rest are filled with \textcolor{blue}{$\blacktriangle$}. Both $\tan{\beta}$ and {$T_{e22}$} are sensitive parameters, and $a_{\mu}^{NBL}$ decreases with the increase of $\tan{\beta}$ and increases with the increase of {$T_{e22}$}, which is consistent with the feature of one-dimensional graph.

In Fig.\ref{D4}, we all select $\tan{\beta}$ as the horizontal axis. It can be seen intuitively that $\tan{\beta}$ has a strong influence on $a_{\mu}^{NBL}$. The vertical axis of Fig.\ref{D4}(a) is $M^2_L$, and the vertical axis of Fig.\ref{D4}(b) is {$T_{e22}$}. One shows an increasing trend and the other shows a decreasing trend. We can conclude that all three are sensitive
parameters. When $7<\tan{\beta}<13$, $ 0.1~{\rm TeV^2} < M^2_L < 0.7~{\rm TeV^2}$,{$~1.5~{\rm TeV} < T_{e22} < 2.5~{\rm TeV}$}, the value of $a_{\mu}^{NBL}$ is closest to the range of 1$\sigma$.

\begin{figure}[ht]
\setlength{\unitlength}{5mm}
\centering
\includegraphics[width=3in]{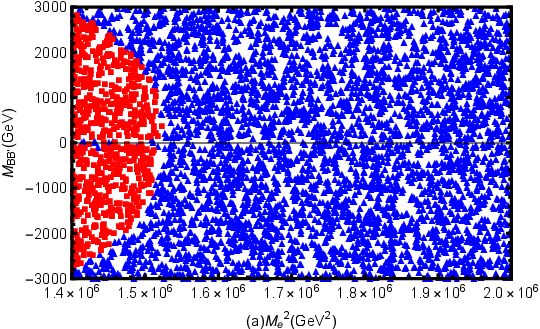}
\vspace{0.2cm}
\setlength{\unitlength}{5mm}
\centering
\includegraphics[width=3.15in]{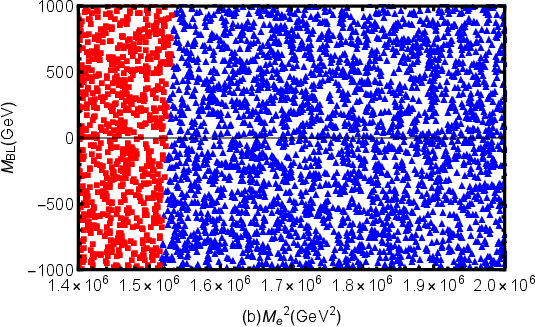}
\caption{$a_{\mu}^{NBL}$ in $M^2_e-M_{BB'}$ plane(a), $M^2_e-M_{BL}$ plane(b).}{\label {D5}}
\end{figure}

{We suppose the parameters with $~T_{e22}=2.4~{\rm TeV}$,$~\tan{\beta}=10$,$~M^2_L=0.16~{\rm TeV^2}$,{$~M_{BL}=1~{\rm TeV}$} in Fig. \ref{D5}(a),
and display a plot of $a_{\mu}^{NBL}$ in the $M^2_e$ versus $M_{BB'}$ plane. The \textcolor{blue}{$\blacktriangle$} occupies most of the space on the right side of the plane, ranging from $1.52~{\rm TeV}^2$ to nearly $2.0~{\rm TeV}^2$ on the horizontal axis. The \textcolor{red}{$\blacksquare$} are located on the left side of the chart, showing a arch, from $1.4~{\rm TeV}^2$ to close to $1.52~{\rm TeV}^2$ on the horizontal axis. $M^2_e$ is an important parameter affecting $a_{\mu}^{NBL}$, and $M^2_e$ appears in the mass matrix of the slepton. The increase of $M^2_e$ makes slepton heavy, which suppresses the contributions from the slepton. Thereby $a_{\mu}^{NBL}$ decreases with the increase of $M^2_e$. The increase in $M_{BB'}$ has affected the neutralino mass matrix, but the effect of $M_{BB'}$ is weak in the figure.}

With {$~T_{e22}=2.4~{\rm TeV}$}, $\tan{\beta}=10$,$~M_{BB'}=0.1~{\rm TeV}$,$~M^2_L=0.16~{\rm TeV^2}$, we plot $a_{\mu}^{NBL}$ in the $M^2_e$ versus $M_{BL}$ plane in the Fig.\ref{D5}(b). The plane is clearly divided into two regions. The left area is \textcolor{red}{$\blacksquare$}, showing that $a_{\mu}^{NBL}$ has a decreasing trend from left to right, and changes significantly with $M^2_e$.  \textcolor{red}{$\blacksquare$} is in the range of {$1.4~{\rm TeV}^2 < M^2_e < 1.52~{\rm TeV}^2$}, \textcolor{blue}{$\blacktriangle$} is in the range of {$1.52~{\rm TeV}^2 < M^2_e < 2~{\rm TeV}^2$}.

{\begin{figure}[ht]
\setlength{\unitlength}{5mm}
\centering
\includegraphics[width=3in]{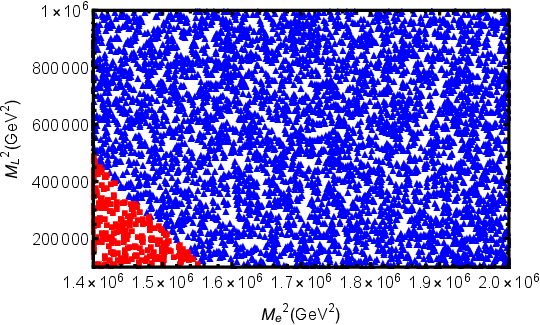}
\caption{$a_{\mu}^{NBL}$ in $M^2_e-M^2_L$ plane.}{\label {e5}}
\end{figure}
}
With {$~T_{e22}=2.4~{\rm TeV}$}, $\tan{\beta}=10$,$~M_{BB'}=0.1~{\rm TeV}$,$~M_{BL}=1~{\rm TeV}$, we plot $a_{\mu}^{NBL}$ in the $M^2_e$ versus $M^2_L$ plane in the Fig.\ref{e5}. Most of the area of the image are \textcolor{blue}{$\blacktriangle$}, and there are \textcolor{red}{$\blacksquare$} in the lower left corner, which is probably distributed in the range of {$1.4~{\rm TeV}^2 < M^2_e < 1.55~{\rm TeV}^2$} and  {$0.1~{\rm TeV}^2 < M^2_L < 0.5~{\rm TeV}^2$}. This is consistent with the trend that $a_{\mu}^{NBL}$ decrease with the increase of $M^2_e$ and $M^2_L$ in the previous images.

\section{Conclusion}
In this paper, we use the effective Lagrangian method to research the one-loop diagrams and some important two-loop diagrams. The studied contributions are composed of one-loop diagrams, the two-loop Barr-Zee type diagrams with fermion sub-loop, the two-loop rainbow type diagrams with fermion sub-loop and the vector bosons($\gamma$, Z, W), the diamond type diagrams in Refs.\cite{two2,our} possessing large factors
\begin{eqnarray}
&& a_\mu=a_{\mu}^{1L}+a_\mu^{2L,~BZ}+a_\mu^{2L,~RB}+a_\mu^{2L,~DIA}.
\end{eqnarray}

We consider the latest experimental constraints and adjust the sensitive parameters. In the end, we obtain rich numerical results and interesting one-dimensional and multi-dimensional scatter plots.

In the one-dimensional graph, we select $~\tan{\beta},{~T_{e22}},~M^2_L,~M^2_e,~M_{BB'},~g_{YB}$ to study {muon anomalous MDM}. Through the trend of the two lines, we can conclude that these parameters play an important role in $a_{\mu}^{NBL}$, where $M^2_e$ and {$T_{e22}$} have a strong influence on $a_{\mu}^{NBL}$. In the scatter plot, we select six parameters $\tan{\beta},{~T_{e22}},~M^2_L,~M^2_e,~M_{BL}$ and {$~M_{BB'}$}. The characteristics of scatter plot is consistent with the one-dimensional diagram, indicating that $M^2_e$ and $\tan{\beta}$ have a strong influence on $a_{\mu}^{NBL}$. Other parameters also have a significant impact, but not as strong as the influence of $M^2_e$ and $\tan{\beta}$ on $a_{\mu}^{NBL}$. From the data obtained in the figure, most numerical results of $a_{\mu}^{NBL}$ are in 2$\sigma$, which can compensate the departure between the experiment data and the SM prediction.
In the used parameter space, the Barr-Zee type two loop contribution to {muon anomalous MDM} is at the order of $10^{-12}\sim 10^{-11}$.
 The rainbow type two loop contribution and the diamond type two loop contribution are both at the order of $10^{-11}\sim 10^{-10}$.
 The ratio of the Barr-Zee type two loop contribution to one loop contribution is around $0.1\%\sim0.4\%$.
 The ratio of the rainbow type two loop contribution to one loop contribution is around $0.5\%\sim6\%$.
 The condition of the diamond type two loop contribution is similar as that of the rainbow type two loop contribution.
Utilizing the parameter space that we have derived to guide experimental design can help optimize experimental conditions
and increase the likelihood of discovering SUSY particles.
After collecting experimental data, we search for data that is consistent with theoretical predictions. By comparing the experimental data with theoretical predictions, we can further refine the parameter space, thereby enhancing the possibility of detecting SUSY particles.

It is well known that there are many  two-loop diagrams that contribute to muon anomalous MDM. Some two-loop diagrams that have not been studied can also give important corrections to muon anomalous MDM, which can further improve the theoretical value. Since the calculation of  two-loop diagrams are very complicated, we will study other  two-loop diagrams in future work.

\begin{acknowledgments}

This work is supported by National Natural Science Foundation of China (NNSFC)
(No.12075074), Natural Science Foundation of Hebei Province
(A2023201040, A2022201022, A2022201017, A2023201041), Natural Science Foundation of Hebei Education Department (QN2022173), Post-graduate's Innovation Fund Project of Hebei University (HBU2024SS042), the Project of the China Scholarship Council (CSC) No. 202408130113.
\end{acknowledgments}

\appendix
\section{Mass matrix and coupling  in N-B-LSSM}\label{A1}

The mass matrix for chargino is:
\begin{eqnarray}
m_{\chi^-} = \left(
\begin{array}{cc}
M_2&\frac{1}{\sqrt{2}}g_2v_\mu\\
\frac{1}{\sqrt{2}}g_2v_d&\frac{1}{\sqrt{2}}\lambda v_S\end{array}
\right).\label{Y2}
 \end{eqnarray}

This matrix is diagonalized by U and V:\begin{eqnarray}
U^{*} m_{\chi^-} V^{\dagger}= m_{\chi^-}^{dia}.
 \end{eqnarray}

The mass matrix for neutrino is:
\begin{eqnarray}
m_{\nu} = \left(
\begin{array}{cc}
0&\frac{1}{\sqrt{2}}v_uY^T_\nu\\
\frac{1}{\sqrt{2}}v_uY_\nu&\ \sqrt{2}v_{\eta}Y_\chi\end{array}
\right).\label{Y2}
\end{eqnarray}

This matrix is diagonalized by $U^V$:
\begin{eqnarray}
U^{V,*}m_\nu U^{V,\dag} = m_\nu^{dia}.
\end{eqnarray}

\begin{equation}
m^2_{A^0} = \left(
\begin{array}{ccccc}
m_{\sigma_{d}\sigma_{d}} &m_{\sigma_{u}\sigma_{d}} &m_{\sigma_1\sigma_{d}} &m_{\sigma_2\sigma_{d}} &m_{{\sigma}_{s}\sigma_{d}}\\
m_{\sigma_{d}\sigma_{u}} &m_{\sigma_{u}\sigma_{u}} &m_{\sigma_1\sigma_{u}} &m_{\sigma_2\sigma_{u}} &m_{{\sigma}_{s}\sigma_{u}}\\
m_{\sigma_{d}\sigma_1} &m_{\sigma_{u}\sigma_1} &m_{\sigma_1\sigma_1} &m_{\sigma_2\sigma_1} &m_{{\sigma}_{s}\sigma_1}\\
m_{\sigma_{d}\sigma_2} &m_{\sigma_{u}\sigma_2} &m_{\sigma_1\sigma_2} &m_{\sigma_2\sigma_2} &m_{{\sigma}_{s}\sigma_2}\\
m_{\sigma_{d}{\sigma}_{s}} &m_{\sigma_{u}{\sigma}_{s}} &m_{\sigma_1{\sigma}_{s}} &m_{\sigma_2{\sigma}_{s}} &m_{{\sigma}_{s}{\sigma}_{s}}\end{array}
\right).\label{CPoddhiggs}
 \end{equation}
Eq.(\ref{CPoddhiggs}) is the CP-odd Higgs mass squared matrix, whose elements are
{\small\begin{eqnarray}
&&m_{\sigma_{d}\sigma_{d}} = m_{H_d}^2+\frac{1}{8} \Big( (g_{1}^{2}+g_{2}^{2}+g_{YB}^{2}) ( v_{d}^{2}  - v_{u}^{2})
 +2 g_{YB} g_{B} ( v_{\eta}^{2}- v_{\bar{\eta}}^{2})\Big)
  + \frac{1}{2}(v_{u}^{2} + v_S^{2})|{\lambda}|^2 ,\nonumber\\&&
m_{\sigma_{d}\sigma_{u}} =\frac{1}{\sqrt{2}} v_S T_{\lambda} +( \frac{1}{2}\kappa v_S^{2}
- \frac{1}{2}{\lambda}_2 v_{\eta} v_{\bar{\eta}}) {\lambda}
, \nonumber\\
&&m_{\sigma_{u}\sigma_{u}} = m_{H_u}^2+\frac{1}{8} \Big( (g_{1}^{2}+g_{YB}^{2}+g_2^2) ( v_{u}^{2}-v_{d}^{2})
 +2 g_{Y B} g_{B} (v_{\bar{\eta}}^{2}- v_{\eta}^{2})\Big)
  + \frac{1}{2}(v_{d}^{2} + v_S^{2})|{\lambda}|^2, \nonumber\\
&&m_{\sigma_1\sigma_1} = m_{\eta}^2+\frac{1}{4} \Big(g_{Y B} g_{B}
 ( v_{d}^{2}- v_{u}^{2})+2g_{B}^{2}(v_{\eta}^{2}- v_{\bar{\eta}}^{2})\Big)+\frac{1}{2} (v_{\bar{\eta}}^{2} + v_S^{2})|{\lambda}_2|^2,\nonumber \\&&
m_{\sigma_1\sigma_2} = \frac{1}{2} \Big(- v_d v_u{\lambda}{\lambda}_{2}
 +  v_S ( \sqrt{2}T_2  +  v_S {\lambda}_{2} \kappa) \Big),\nonumber
\\&&m_{\sigma_2\sigma_2} = m_{\bar{\eta}}^2
 +\frac{1}{4} \Big(g_{Y B} g_{B}
 ( v_{u}^{2}- v_{d}^{2})-2g_{B}^{2} (v_{\eta}^{2}-v_{\bar{\eta}}^{2} )\Big)+\frac{1}{2} (v_{\eta}^{2} + v_S^{2})|{\lambda}_{2}|^2, \nonumber\\
&& m_{\sigma_{d}{\sigma}_{s}} = - v_u \Big(v_S\kappa
{\lambda}  -\frac{1}{\sqrt{2}}T_{{\lambda}} \Big),~~~~~~~~~~~m_{\sigma_{d}\sigma_1} = \frac{1}{2} v_u v_{\bar{\eta}} {\lambda}_2 {\lambda}, \nonumber\\&&
m_{\sigma_{u}{\sigma}_{s}} = - v_d \Big(v_S\kappa
{\lambda}  -\frac{1}{\sqrt{2}}T_{{\lambda}} \Big),~~~~~~~~~~~m_{\sigma_{u}\sigma_1} =\frac{1}{2} v_d v_{\bar{\eta}} {\lambda}_2 {\lambda},\nonumber\\&&
m_{\sigma_1{\sigma}_{s}} =- v_{\bar{\eta}} \Big(v_S\kappa
{\lambda}_2  -\frac{1}{\sqrt{2}}T_2 \Big),~~~~~~~~~~m_{\sigma_{d}\sigma_2} =\frac{1}{2} v_u v_{\eta} {\lambda}_2 {\lambda},\nonumber\\&&
m_{\sigma_2{\sigma}_{s}} = - v_{\eta} \Big(v_S\kappa
{\lambda}_2  -\frac{1}{\sqrt{2}}T_2) \Big),~~~~~~~~~m_{\sigma_{u}\sigma_2} =\frac{1}{2} v_d v_{\eta} {\lambda}_2 {\lambda},\nonumber\\&&
m_{{\sigma}_{s}{\sigma}_{s}} = m^2_{S}+(\kappa v_S^{2}+ {\lambda}_{2} v_{\eta} v_{\bar{\eta}}
+ {\lambda} v_d v_u )\kappa +\frac{1}{2}|{\lambda}|^2 (v_d^2+ v_u^2)
 + \frac{1}{2}|{\lambda}_{2}|^2 (v_{\eta}^2+v_{\bar{\eta}}^2)
\nonumber\\&&\hspace{1.5cm}  -\sqrt{2} v_S T_{\kappa}  .
\end{eqnarray}}

In the basis $(\tilde{e}_L, \tilde{e}_R)$, the mass matrix for slepton is shown and diagonalized by $Z^E$ through the
formula $Z^E m^2_{\tilde{e}} Z^{E,\dagger} = m^{diag}_{2,\tilde{e}}$,
\begin{equation}
m^2_{\tilde{e}} = \left(
\begin{array}{cc}
m_{\tilde{e}_L\tilde{e}_L^*} &\frac{1}{\sqrt{2}}  v_d T_{e}^{\dagger}  -\frac{1}{2} v_u {\lambda} v_s Y_{e}^{\dagger} \\
\frac{1}{\sqrt{2}} \sqrt{2} v_d T_{e}  - \frac{1}{2}v_u {\lambda^*} v_s Y_{e}  &m_{\tilde{e}_R\tilde{e}_R^*}\end{array}
\right),
 \end{equation}
\begin{eqnarray}
&&m_{\tilde{e}_L\tilde{e}_L^*} = m_{\tilde{L}}^2+\frac{1}{8} \Big((g_{1}^{2} + g_{Y B}^{2}
+ g_{Y B} g_{B} -g_2^2)(v_{d}^{2}- v_{u}^{2})+ 2(g_{B}^2+ g_{Y B} g_{B})( v_{\eta}^{2}- v_{\bar{\eta}}^{2}
)
\Big)+\frac{v_{d}^{2}}{2} {Y_{e}^2} ,\nonumber\\&&
m_{\tilde{e}_R\tilde{e}_R^*} = m_{\tilde{E}}^2-\frac{1}{8}  \Big([2(g_{1}^{2} + g_{Y B}^{2})+g_{Y B} g_{B}]
( v_{d}^{2}- v_{u}^{2})\nonumber\\&&\hspace{1.7cm}+(4g_{Y X} g_{B}+2g_{B}^{2})(v_{\eta}^{2}- v_{\bar{\eta}}^{2})
\Big)+\frac{1}{2} v_{d}^{2} {  Y_{e}^2}.
\end{eqnarray}

The used vertexes in Eq.(\ref{OL}) are:
\begin{eqnarray}
&&\hspace{0cm}A_R=\frac{1}{\sqrt{2}}Z_{k2}^{E}(g_1N_{j1}^{*}+g_2N_{j2}^{*}+g_{YB}N_{j5}^{*})
-N_{j3}^{*}Y_\mu Z_{k5}^{E},
\nonumber\\&&\hspace{0cm}
A_L=-\frac{1}{\sqrt{2}}Z_{k5}^{E}[2g_1N_{j1}+(2g_{YB}+g_B)N_{j5}]-Y_\mu^{*}Z_{k2}^EN_{j3},
\nonumber\\&&\hspace{0cm}B_L=-\frac{1}{\sqrt{2}}U_{i2}^{*}Z_{k2}^{I*}Y_\mu,~~~~~~~~~~
B_R=\frac{1}{\sqrt{2}}g_2Z_{k2}^{I*}V_{i1},
\nonumber\\&&\hspace{0cm}C_L=\frac{1}{\sqrt{2}}U_{i2}^{*}Z_{k2}^{R*}Y_\mu,~~~~~~~~~~~~
C_R=-\frac{1}{\sqrt{2}}g_2Z_{k2}^{R*}V_{i1}.
\end{eqnarray}

\begin{eqnarray}
M_{\nu}=
\left({\begin{array}{*{20}{c}}
0 & \frac{\upsilon_u}{\sqrt{2}}(Y_\nu^T)^{IJ}  \\
\frac{\upsilon_u}{\sqrt{2}}(Y_\nu)^{IJ} & \sqrt{2}\upsilon_{\bar{\eta}}(Y_\chi)^{IJ}  \\
\end{array}}
\right),~~~~~~~ {\rm with}~~~ I,J=1,2,3.
\end{eqnarray}
The effective light neutrino mass matrix is in general given as $m_{eff}=-mM^{-1}m^{T}$,
with
\begin{eqnarray}
&&m=\frac{1}{\sqrt{2}}v_uY^T_\nu,~~~~~M=\sqrt{2}v_{\overline{\eta}}Y_\chi,\nonumber\\
\end{eqnarray}

{\section{The one loop results in MIA}
Using mass insertion approximation method,
we calculate the one loop contribution to {muon anomalous MDM} in the N-B-LSSM.}

 {1. Chargino and sneutrino(CP-even and CP-odd) contributions
\begin{eqnarray}
&&a_\mu(\tilde{\nu}^R_L, \tilde{H}^-, \tilde{W}^-)
=\frac{g_2^2}{2}
x_\mu\sqrt{x_2x_{\mathcal{H}}}\tan\beta[2\mathcal{I}_1(x_{\mathcal{H}},x_{\tilde{\nu}^R_L},x_2)
-\mathcal{I}_2(x_2,x_{\mathcal{H}},x_{\tilde{\nu}^R_L})]\label{MIARC},
\\&&a_\mu(\tilde{\nu}^I_L, \tilde{H}^-, \tilde{W}^-)
=\frac{g_2^2}{2}
x_\mu\sqrt{x_2x_{\mathcal{H}}}\tan\beta[2\mathcal{I}_1(x_{\mathcal{H}},x_{\tilde{\nu}^R_L},x_2)
-\mathcal{I}_2(x_2,x_{\mathcal{H}},x_{\tilde{\nu}^R_L})]\label{MIAIC}.
\end{eqnarray}
Here $m_{\mathcal{H}}=\frac{\lambda_H v_S}{\sqrt{2}}$
and $x_{\mathcal{H}}=\frac{m_{\mathcal{H}}^2}{\Lambda^2}$.}

 {The one-loop functions $\mathcal{I}_1(x,y,z)$ and $\mathcal{I}_2(x,y,z)$ are defined as
\begin{eqnarray}
&&\mathcal{I}_1(x,y,z)=\frac{y-x}{(x-y)^2 (y-z)}+\frac{y \log x}{(x-y)^2
   (x-z)}\nonumber\\&&\hspace{1.8cm}+\frac{y (x-2 y+z)\log y}{(x-y)^2
   (y-z)^2}-\frac{y \log z}{(x-z) (y-z)^2},\\&&
\mathcal{I}_2(x,y,z)=\frac{2 z [z^3-3 x y z+x y
   (x+y)]\log z}{(x-z)^3 (y-z)^3}+\frac{2 x z \log x}{(x-y)
   (x-z)^3}\nonumber\\&&\hspace{1.8cm}+\frac{x (y+z)+z (y-3
   z)}{(x-z)^2 (y-z)^2}-\frac{2 y z \log y}{(x-y) (y-z)^3}.
\end{eqnarray}}

 {2. The one-loop contributions from $\tilde{B}(\tilde{B}^\prime$)-$\tilde{\mu}_L$-$\tilde{\mu}_R$.
\begin{eqnarray}
&&a_\mu(\tilde{\mu}_R,\tilde{\mu}_L, \tilde{B})=
g_1^2x_\mu\sqrt{x_1x_{\mathcal{H}}}
\tan\beta~\mathcal{I}_3(x_1,x_{\tilde{\mu}_L},x_{\tilde{\mu}_R})\label{MIABLR}
,\\
&&a_\mu(\tilde{\mu}_R,\tilde{\mu}_L, \tilde{B}^\prime)
=(g_{YB}+\frac{g_B}{2})(g_{YB}+g_B)x_\mu\sqrt{x_{\tilde{B}^\prime} x_{\mathcal{H}}}\tan\beta
\mathcal{I}_3~(x_{\tilde{B}^\prime},x_{\tilde{\mu}_L},x_{\tilde{\mu}_R})\label{MIAXLR}.
\end{eqnarray}
The one-loop function $\mathcal{I}_3(x,y,z)$ is
\begin{eqnarray}
&&\mathcal{I}_3(x,y,z)=\frac{2 x[x^3-3 x y z+y z
   (y+z)] \log x}{(x-y)^3 (x-z)^3}-\frac{2 x y \log y}{(x-y)^3
   (y-z)}\nonumber\\&&\hspace{1.8cm}+\frac{x (y+z)-3 x^2+y
   z}{(x-y)^2 (x-z)^2}+\frac{2 x z \log z}{(x-z)^3 (y-z)}.
\end{eqnarray}}

 {3. The one-loop contributions from $\tilde{B}({\tilde{B}^\prime})$-$\tilde{H}^0$-$\tilde{\mu}_R$.
\begin{eqnarray}
&&a_\mu(\tilde{\mu}_R, \tilde{B}, \tilde{H}^0)
=-g_1^2x_\mu\sqrt{x_1x_{\mathcal{H}}}\tan\beta~
\mathcal{I}_2(x_1,x_{\mathcal{H}},x_{\tilde{\mu}_R})\label{MIAHBR},
\\&&a_\mu(\tilde{\mu}_R, \tilde{B}^\prime, \tilde{H}^0)
=-g_{YB}(g_{YB}+\frac{g_B}{2})x_\mu\sqrt{x_{\tilde{B}^\prime}x_{\mathcal{H}}}\tan\beta~
\mathcal{I}_2(x_{\tilde{B}^\prime},x_{\mathcal{H}},x_{\tilde{\mu}_R})\label{MIAHXR}.
\end{eqnarray}}

 {4. The one-loop contributions from $\tilde{B}(\tilde{W}^0,{\tilde{B}^\prime})$-$\tilde{H}^0$-$\tilde{\mu}_L$.
\begin{eqnarray}
&&a_\mu(\tilde{\mu}_L, \tilde{H}^0, \tilde{B})
=\frac{1}{2}g_1^2x_\mu\sqrt{x_1x_{\mathcal{H}}}\tan\beta~
\mathcal{I}_2(x_1,x_{\mathcal{H}},x_{\tilde{\mu}_L})\label{MIABHL},
\\&&a_\mu(\tilde{\mu}_L, \tilde{H}^0, \tilde{W}^0)=-\frac{1}{2}g_2^2
x_\mu\sqrt{x_2x_{\mathcal{H}}}\tan\beta~
\mathcal{I}_2(x_2,x_{\mathcal{H}},x_{\tilde{\mu}_L})\label{MIAWHL},
\\&&a_\mu(\tilde{\mu}_L, \tilde{H}^0, {\tilde{B}^\prime})
=\frac{1}{2}g_{YB}(g_{YB}+g_B)x_\mu\sqrt{x_{\tilde{B}^\prime}x_{\mathcal{H}}}\tan\beta
~\mathcal{I}_2(x_{\tilde{B}^\prime},x_{\mathcal{H}},x_{\tilde{\mu}_L})\label{MIAXHL}.
\end{eqnarray}}

 {5. The one-loop contributions from $\tilde{B}-{\tilde{B}^\prime}-\tilde{\mu}_R-\tilde{\mu}_L$.
\begin{eqnarray}
&&a_\mu(\tilde{\mu}_R,\tilde{\mu}_L, \tilde{B}, {\tilde{B}^\prime})
=g_1(4g_{YB}+3g_B)x_\mu \sqrt{x_{BB^\prime}x_{\mathcal{H}}}\tan\beta
\nonumber\\&&\times\Big(\sqrt{x_1x_{\tilde{B}^\prime}}
f(x_{\tilde{B}^\prime},x_1,x_{\tilde{\mu}_L},x_{\tilde{\mu}_R}) -g(x_{\tilde{B}^\prime},x_1,x_{\tilde{\mu}_L},x_{\tilde{\mu}_R})
\Big)\label{MIAXBLR}.
\end{eqnarray}
The one loop functions $f(x,y,z,t)$ and $g(x,y,z,t)$ are shown as
\begin{eqnarray}
&&f(x,y,z,t)=
   \frac{1}{16\pi^2}\Big[\frac{t [t^3-3 t x y+x y
   (x+y)]\log t}{(t-x)^3 (t-y)^3 (t-z)}-\frac{x[x^3-3 t x z+t z (t+z)] \log x
}{(t-x)^3 (x-y)
   (x-z)^3}\nonumber\\&&+\frac{y [y^3-3 t y z+t z
   (t+z)]\log y}{(t-y)^3 (x-y) (y-z)^3}-\frac{z[z^3-3 x y z+x y (x+y)] \log z}{(t-z) (z-x)^3
   (z-y)^3}+\frac{1}{2
   (x-y)}\nonumber\\&&\times\Big(\frac{t}{(t-x)^2 (z-x)}-\frac{2y}{(t-y) (y-z)^2}+\frac{x (2 t-3
   x+z)}{(t-x)^2 (x-z)^2}+\frac{t+y}{(t-y)^2 (y-z)}\Big)\Big],
\\&&g(x,y,z,t)=\frac{1}{16\pi^2}\Big\{-\frac{t [t^3 (x+y)-3 t^2 x y+x^2
   y^2]\log t }{(t-x)^3 (t-y)^3 (t-z)}\nonumber\\&&+\frac{z [x^2 y^2+x z^2 (z-3 y)+y z^3]\log z
   }{(t-z) (z-x)^3
   (z-y)^3}+\frac{x^2[x^3-3 t x z+t z
   (t+z)] \log x}{(t-x)^3 (x-y) (x-z)^3}\nonumber\\&&-\frac{y^2 [y^3-3 t y z+t z (t+z)]\log y }{(t-y)^3 (x-y)
   (y-z)^3}-\frac{x^2 (2 t-3
   x+z)}{2(t-x)^2 (x-y) (x-z)^2}\nonumber\\&&+\frac{t x}{2(t-x)^2 (x-y)
   (x-z)}-\frac{y [t (y+z)+y (z-3 y)]}{2(t-y)^2 (y-x)
   (y-z)^2}\Big\}.
\end{eqnarray}}

 {In Eqs.(\ref{MIARC}), (\ref{MIAIC}), (\ref{MIABLR}), (\ref{MIAXLR}), (\ref{MIAHBR}) $\dots$ (\ref{MIAXBLR}),
 one can easily find the factor $x_\mu\tan\beta$ with $x_\mu=\frac{m_\mu^2}{\Lambda^2}$, which is similar as the MSSM condition.
 Eqs.(\ref{MIAXLR}), (\ref{MIAHXR}), (\ref{MIAXHL}), (\ref{MIAXBLR}),
  include the new gauge coupling constants $g_B$ and $g_{YB}$, which are beyond MSSM.}

 {Supposing  all the masses of the superparticles are almost degenerate.
 we also use the following relation to obtain simplified results
\[M_1=M_2=m_{\mathcal{H}}=m_{\tilde{\mu}_L}
=m_{\tilde{\mu}_R}=m_{\tilde{\nu}^{R,I}_L}
=m_{\tilde{\nu}_R}^{R,I}=|M_{BB^\prime}|=|M_{\tilde{B}^\prime}|=M_{SUSY}.\]}

 {Then these one loop functions $\mathcal{I}_1(x,y,z),~\mathcal{I}_2(x,y,z),~\mathcal{I}_3(x,y,z),~f(x,y,z,t),~g(x,y,z,t)$ are much simplified as
\begin{eqnarray}
&&\mathcal{I}_1(1,1,1)= \frac{1}{48\pi^2},~~~~\mathcal{I}_2(1,1,1)=\frac{1}{96\pi^2},~~~~\mathcal{I}_3(1,1,1)=\frac{1}{96\pi^2},
\nonumber\\&&
 f(1,1,1,1)= -\frac{1}{240 \pi^2 },~~~~g(1,1,1,1)= -\frac{1}{960 \pi ^2}.
\end{eqnarray}
Using the relations, N-B-LSSM one-loop contributions to muon g-2 are simplified to a large extent.
\begin{eqnarray}
&&a^{1L}_\mu\simeq \frac{1}{192\pi^2}\frac{m_\mu^2}{M_{SUSY}^2}\tan\beta(5g_2^2+g_1^2)\nonumber\\&&
\nonumber\\&&+\frac{1}{192\pi^2}\frac{m_\mu^2}{M_{SUSY}^2}\tan\beta\texttt{sign}[M_{\tilde{B}^\prime}](g_B^2+3g_{YB}g_B+g_{YB}^2)
\nonumber\\&&+\frac{1}{960\pi^2}\frac{m_\mu^2}{M_{SUSY}^2}\tan\beta
g_1(4g_{YB}+3g_B)\texttt{sign}[M_{BB^\prime}]
\Big(1-4\texttt{sign}[M_{\tilde{B}^\prime}]\Big).\label{amuS}
\end{eqnarray}}

\end{document}